**Preserving EP-HistARy through Augmented Reality**

Preserving EP-HistARy is an extended reality prototype that is developed from Emile Pitre's story as one of the founding members of the Black Student Union at the University of Washington.

Annie, AY, Yang

University of Washington, anyang95@uw.edu

**ABSTRACT**

Extended reality as a technology can weave together the fabric of the past, present, and future. Through Yolanda Barton's Revere XR startup, a 2-day design hackathon was held to bring the community together through a love for history and using technology for good. Through interviewing an influential community elder, Emile Pitre, and referencing his book "Revolution to Evolution", my team developed an augmented reality artifact to tell his story and preserve one revolutionary man's legacy that impacted the University of Washington's history forever.

**CCS CONCEPTS**

• Human-centered computing

## 1 INTRODUCTION

Revere XR, a mixed-reality startup, hosted a two-day design hackathon at the University of Washington to facilitate community passion and engagement in preserving Seattle's histories and culture through immersive reality technology. Revere is a verb that means to honor and respect profoundly, a deep appreciation and admiration for those that have paved the way. Embodying the mission in every aspect, Yolanda Barton, the founder, invited a panel of influential elders to speak about their stories of impact growing up. Each team of was comprised of students ranging from undergraduates, master's, and PhD, assuming either the function of "Storyteller", "Community", "Tech", or "Design" to encompass all aspects of effective problem-solving and storytelling. Each team was also introduced to, and assigned an elder to interview and learn from to help guide the ideation and design process of the augmented reality or virtual reality prototype. My team interviewed Emile Pitre, one of the founding Black Student Union members at the University of Washington, who paved the way for minorities to attend the university in the 1960s and directly influenced the emergence of the Office of Minority Affairs and Diversity.

## 2 BACKGROUND RESEARCH

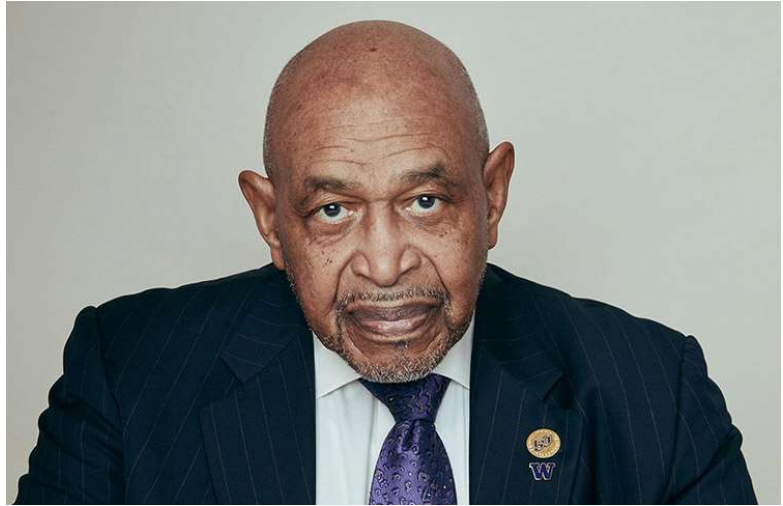

Figure 1: 2021 A portrait of Emile Pitre. Photograph by Quinn Russel Brown, via Seattle Medium.
(https://seattlemedium.com/emile-pitre-the-ongoing-battle-for-diversity-at-uw/ )

Emile Pitre is a lifelong activist, author, and former University of Washington graduate student who participated in the protests in the late 1960s, demanding for UW President Charles Odegaard and the institution to allow marginalized groups of color to obtain the same right to education as their white counterparts on campus. As a team, we were honored to have given a chance to interview him about his childhood and youth leading up to the early defining moments of his career and legacy. He was the first in his family to graduate high school, earning a full ride scholarship to Southern University in Baton Rouge, before attending University of Washington as a graduate student in chemistry in 1967.Then, in 1968, he, along with his founding Black Student Union members and allies, occupied Charles Odegaard's office to negotiate the BSU's demands for more diversity and equity at the university. After a period of negotiation, Odegaard established what is now known as the Office of Minority Affairs & Diversity (OMA&D), the first of its kind at any university in the United States. He authored the book "Revolution to Evolution", which accounts the details of chronological ordered events that directly impacted the emergence of OMA&D. Pitre is also famous for his efforts and contribution given to the Instructional Center, a program established through the OMA&D that offers academic tutoring to students.

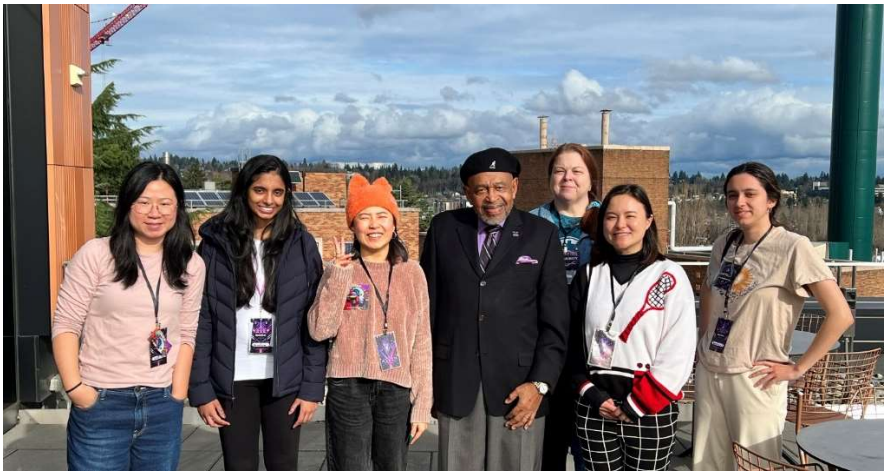

Figure 2: 2024 Team Quantum SpaRX at Revere XR Hackathon, post-interview with Emile Pitre.

# 3 THEMATIC ANALYSIS & IDEATION

To strengthen our research and support our ideation process, we borrowed the "Revolution to Evolution" book from the campus library. Using this to cross-reference and confirm our understanding of historical events helped with approaching ideation with an informed lens. It was important to not rush ahead and design a solution without considering all the knowledge we have gathered from the interview, and his physical book. We pulled out key dates from the book to outline the chronological timeline of Pitre's life to figure out how to form our story. Through affinity diagramming, we gathered that the key dates and defining moments to visualize the timeline and craft the storytelling around involved the years 1967, when Pitre became a graduate student at the University of Washington, and 1968, the year of protests and demands urging the school to allow not only Black students, but also Mexicans, Indigenous, Asian, and other minority students to come to campus, that lead to the founding of the OMA&D. We also included 1970, the year Vice President Samuel E Kelly founded the Ethnic Culture Center (part of OMA&D).

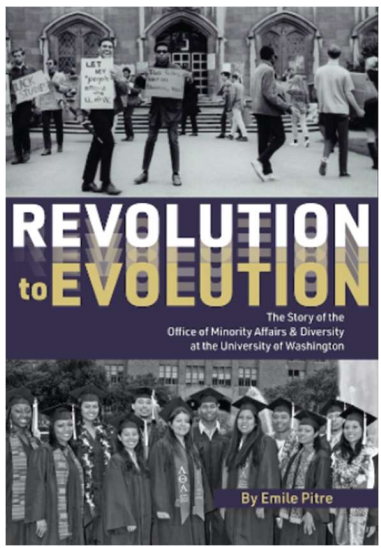
Figure 3.1: The cover of Emile Pitre's book "Revolution to Evolution".

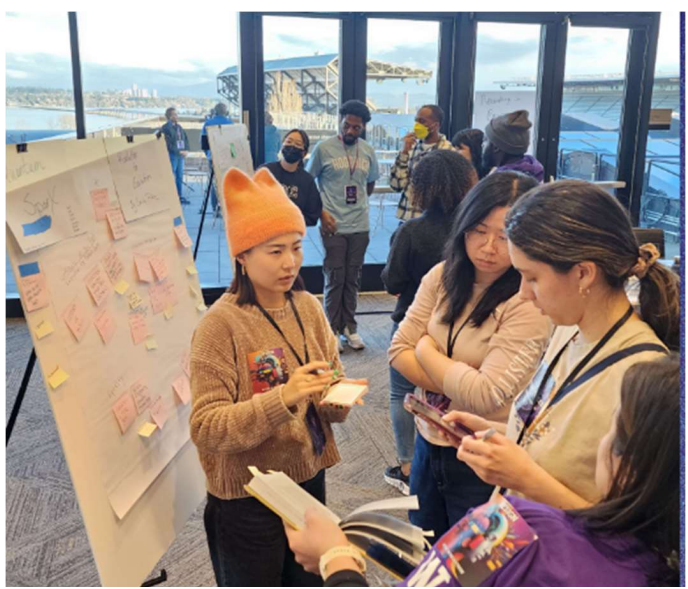
Figure 3.2: 2024 Team Quantum SpaRX at Revere XR Hackathon running an affinity diagramming session and pulling dates from Emile Pitre's

book, "Revolution to Evolution".

It was a team goal for us to approach this project **to instill** an emotional connection to our audience, so in seeking feedback from hackathon tech experts, we chose to narrow down on one key defining moment in 1968 instead of trying to write a story around multiple dates. This choice guided our storyboarding, where we ideated scenarios to emphasize the historical impact of Pitre's revolutionary actions. Most students on campus do not receive an active opportunity to learn about the racial history of **the** University of Washington's educational programming. We chose to focus on immersing the audience in the iconic sit-in of 1968, **which** added pressure to Odegaard's decision **to let** minority student groups be educated at the institution.

## 4  PROTOTYPE

Initially on day one, we tinkered and experimented with Unity to explore whether virtual reality would be a viable medium for developing the final prototype. The Techs and Storyteller worked together to tackle the learning curve so we don't get stuck on day two. However, virtual reality is not accessible to share the outcome and impact to a wide audience, as the physical wearable element confines the user within the headset environment. Therefore, we pivoted to augmented reality and learned to use Mixamo to change the seating and standing positions of the avatars, Adobe Aero to integrate the avatars and the background setting, and Blender to weave all the augmented elements together for the final Figma prototype demonstration.

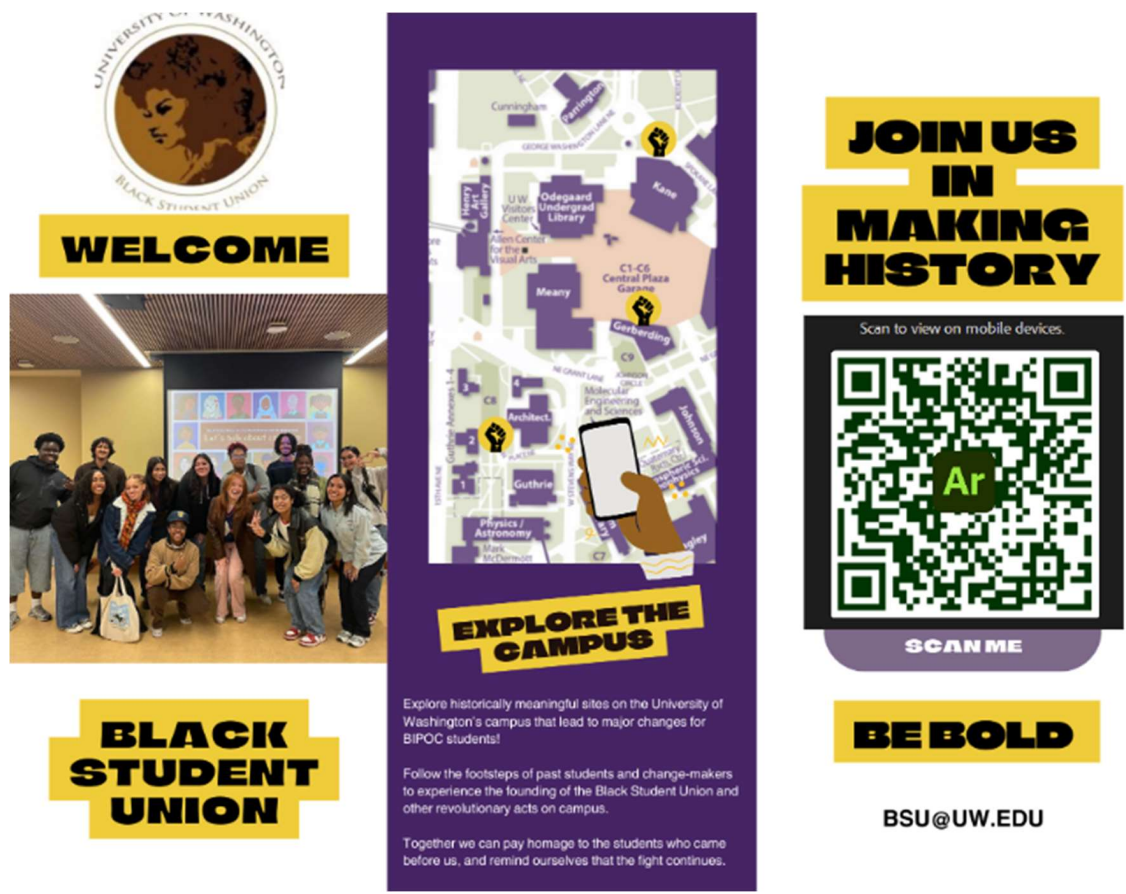

Figure 4.1: Digital student brochure with QR code to access augmented reality experience.

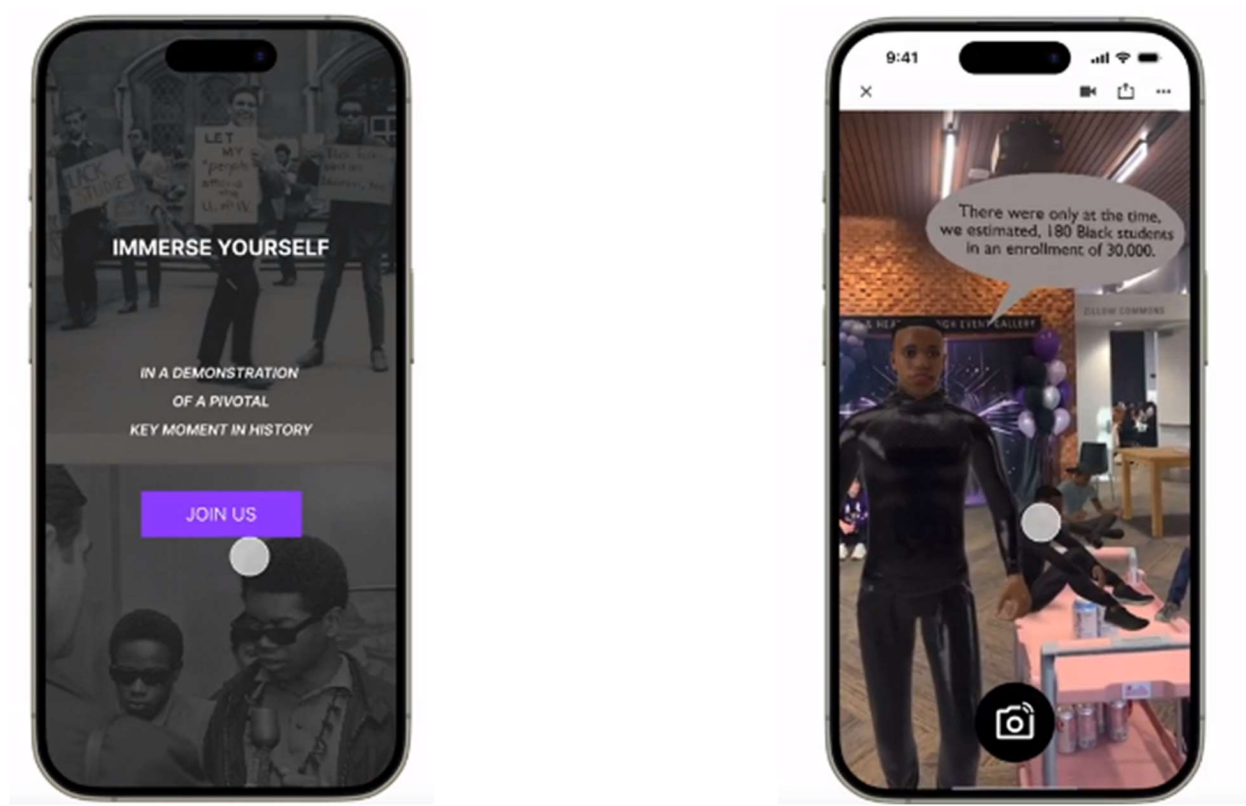

Figure 4.2: Example screens of the final prototyped artifact made using Mixamo, Blender, and Adobe Aero.

Imagine that as a new student on campus, your orientation leader is showing you around the campus and speaking about history in the making of organizations like the Black Student Union. You are handed a brochure to learn more about the Black Student Union, and open it up to find a QR code, enticing you to pull out your phone and see what it will show you. The QR code leads you to a demonstration of an immersive sit-in that Emile Pitre participated in, and you learn about the history and legacy in an engaging, interactive, and memorable activity. We chose to design the interaction as an interactive historical timeline walkthrough so new students can immerse themselves in each key moment of the campus history. After the period is chosen, the immersion can be accessed to re-enact the scene through augmented reality. This meets Pitre's desire to continue to involve the modern student body to join in on the continued journey to freedom and securing human rights for underserved communities.

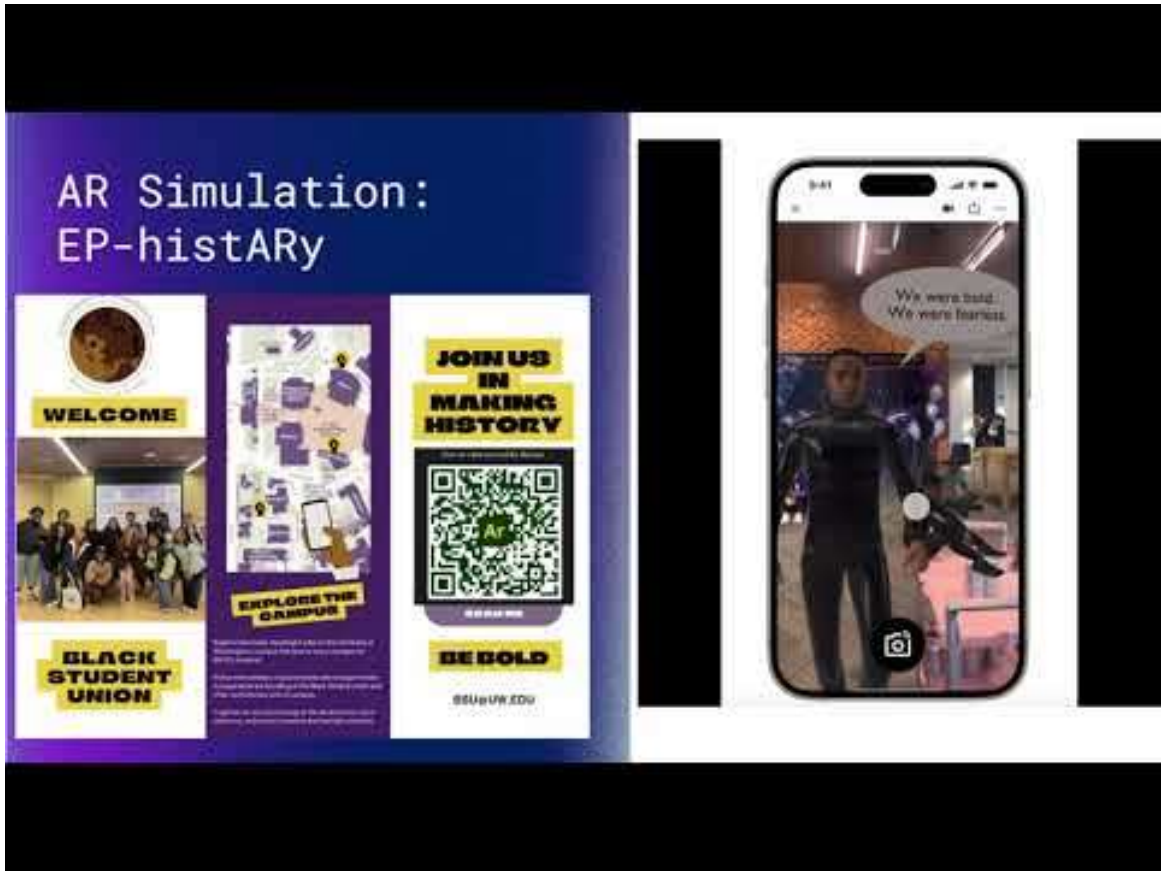

https://youtu.be/1KcDAxcg9Ew

**ACKNOWLEDGMENTS**

I'd like to thank Yolanda Barton and her Revere XR team, the partners, sponsors, tech experts, and XR developers from Niantic, Nvidia, mixed reality leaders, and the judging panel comprising of Julia C. Beabout, Ana Pinto de Silva, Erin Reddick, Chang Liu, and Chloe Fulton**.** For a diverse, innovative, and engaging design hackathon and learning experience. To Rafael Silva and Dayton Kelly for informing me of this opportunity. To Melissa Ewing, and the rest of the planning committee at Paul G. Allen School of Computer Science and Human-Centered Design and Engineering programs.